
\newcount\mgnf\newcount\tipi\newcount\tipoformule\newcount\driver \driver=1
\mgnf=0          
\tipi=2          
\tipoformule=1   
\ifnum\mgnf=0
   \magnification=\magstep0\hoffset=0.cm
   \voffset=-0.5truecm\hsize=15.5truecm\vsize=22.truecm
   \parindent=4.pt\fi
\ifnum\mgnf=1
   \magnification=\magstep1\hoffset=0.truecm
   \voffset=-0.5truecm\hsize=15.5truecm\vsize=22.truecm
   \baselineskip=14pt plus0.1pt minus0.1pt \parindent=12pt
   \lineskip=4pt\lineskiplimit=0.1pt      \parskip=0.1pt plus1pt\fi
%
%
\let\a=\alpha       \let\d=\delta \let\e=\varepsilon
     \let\th=\vartheta \let\l=\lambda
    \let\n=\nu               
\let\s=\sigma \let\t=\tau    \let\f=\varphi
  \let\o=\omega


{\count255=\time\divide\count255 by 60 \xdef\oramin{\number\count255}
        \multiply\count255 by-60\advance\count255 by\time
   \xdef\oramin{\oramin:\ifnum\count255<10 0\fi\the\count255}}

\def\ora{\oramin }

\def\data{\number\day/\ifcase\month\or gennaio \or febbraio \or marzo \or
aprile \or maggio \or giugno \or luglio \or agosto \or settembre
\or ottobre \or novembre \or dicembre \fi/\number\year;\ \ora}

\setbox200\hbox{$\scriptscriptstyle \data $}

\newcount\pgn \pgn=1
\def\foglio{\number\numsec:\number\pgn
\global\advance\pgn by 1}
\def\foglioa{A\number\numsec:\number\pgn
\global\advance\pgn by 1}

%


\global\newcount\numsec\global\newcount\numfor
\global\newcount\numfig
\gdef\profonditastruttura{\dp\strutbox}
\def\senondefinito#1{\expandafter\ifx\csname#1\endcsname\relax}
\def\SIA #1,#2,#3 {\senondefinito{#1#2}
\expandafter\xdef\csname #1#2\endcsname{#3} \else
\write16{???? ma #1,#2 e' gia' stato definito !!!!} \fi}
\def\etichetta(#1){(\veroparagrafo.\veraformula)
\SIA e,#1,(\veroparagrafo.\veraformula)
 \global\advance\numfor by 1
 \write16{ EQ \equ(#1) == #1  }}
\def \FU(#1)#2{\SIA fu,#1,#2 }
\def\etichettaa(#1){(A\veroparagrafo.\veraformula)
 \SIA e,#1,(A\veroparagrafo.\veraformula)
 \global\advance\numfor by 1
 \write16{ EQ \equ(#1) == #1  }}
\def\getichetta(#1){Fig. \verafigura
 \SIA e,#1,{\verafigura}
 \global\advance\numfig by 1
 \write16{ Fig. \equ(#1) ha simbolo  #1  }}
\newdimen\gwidth
\def\BOZZA{
\def\alato(##1){
 {\vtop to \profonditastruttura{\baselineskip
 \profonditastruttura\vss
 \rlap{\kern-\hsize\kern-1.2truecm{$\scriptstyle##1$}}}}}
\def\galato(##1){ \gwidth=\hsize \divide\gwidth by 2
 {\vtop to \profonditastruttura{\baselineskip
 \profonditastruttura\vss
 \rlap{\kern-\gwidth\kern-1.2truecm{$\scriptstyle##1$}}}}}
\footline={\rlap{\hbox{\copy200}\ $\st[\number\pageno]$}\hss\tenrm
\foglio\hss}
}
\def\alato(#1){}
\def\galato(#1){}
\def\veroparagrafo{\number\numsec}\def\veraformula{\number\numfor}
\def\verafigura{\number\numfig}
\def\geq(#1){\getichetta(#1)\galato(#1)}
\def\Eq(#1){\eqno{\etichetta(#1)\alato(#1)}}
\def\eq(#1){\etichetta(#1)\alato(#1)}
\def\Eqa(#1){\eqno{\etichettaa(#1)\alato(#1)}}
\def\eqa(#1){\etichettaa(#1)\alato(#1)}
\def\eqv(#1){\senondefinito{fu#1}$\clubsuit$#1\write16{No translation for #1}%
\else\csname fu#1\endcsname\fi}
\def\equ(#1){\senondefinito{e#1}\eqv(#1)\else\csname e#1\endcsname\fi}

\ifnum\tipoformule=1\let\Eq=\eqno\def\eq{}\let\Eqa=\eqno\def\eqa{}
\def\equ(#1){(#1)}\fi

\def\include#1{
\openin13=#1.aux \ifeof13 \relax \else
\input #1.aux \closein13 \fi}
\openin14=\jobname.aux \ifeof14 \relax \else
\input \jobname.aux \closein14 \fi
%

\newskip\ttglue
\def\TIPI{
\font\ottorm=cmr8   \font\ottoi=cmmi8
\font\ottosy=cmsy8  \font\ottobf=cmbx8
\font\ottott=cmtt8  
\font\ottoit=cmti8
\def \ottopunti{\def\rm{\fam0\ottorm}
\textfont0=\ottorm  \textfont1=\ottoi
\textfont2=\ottosy  \textfont3=\ottoit
\textfont4=\ottott
\textfont\itfam=\ottoit  \def\it{\fam\itfam\ottoit}%
\textfont\ttfam=\ottott  \def\tt{\fam\ttfam\ottott}%
\textfont\bffam=\ottobf
\normalbaselineskip=9pt\normalbaselines\rm}
\let\nota=\ottopunti}
\def\TIPIO{
\font\setterm=amr7 
\font\settesy=amsy7 \font\settebf=ambx7 
\def \settepunti{\def\rm{\fam0\setterm}
\textfont0=\setterm   
\textfont2=\settesy   
\textfont\bffam=\settebf  \def\bf{\fam\bffam\settebf}
\normalbaselineskip=9pt\normalbaselines\rm
}\let\nota=\settepunti}

\def\TIPITOT{
\font\twelverm=cmr12
\font\twelvei=cmmi12
\font\twelvesy=cmsy10 scaled\magstep1
\font\twelveex=cmex10 scaled\magstep1
\font\twelveit=cmti12
\font\twelvett=cmtt12
\font\twelvebf=cmbx12
\font\twelvesl=cmsl12
\font\ninerm=cmr9
\font\ninesy=cmsy9
\font\eightrm=cmr8
\font\eighti=cmmi8
\font\eightsy=cmsy8
\font\eightbf=cmbx8
\font\eighttt=cmtt8
\font\eightsl=cmsl8
\font\eightit=cmti8
\font\sixrm=cmr6
\font\sixbf=cmbx6
\font\sixi=cmmi6
\font\sixsy=cmsy6
\font\twelvetruecmr=cmr10 scaled\magstep1
\font\twelvetruecmsy=cmsy10 scaled\magstep1
\font\tentruecmr=cmr10
\font\tentruecmsy=cmsy10
\font\eighttruecmr=cmr8
\font\eighttruecmsy=cmsy8
\font\seventruecmr=cmr7
\font\seventruecmsy=cmsy7
\font\sixtruecmr=cmr6
\font\sixtruecmsy=cmsy6
\font\fivetruecmr=cmr5
\font\fivetruecmsy=cmsy5
\textfont\truecmr=\tentruecmr
\scriptfont\truecmr=\seventruecmr
\scriptscriptfont\truecmr=\fivetruecmr
\textfont\truecmsy=\tentruecmsy
\scriptfont\truecmsy=\seventruecmsy
\scriptscriptfont\truecmr=\fivetruecmr
\scriptscriptfont\truecmsy=\fivetruecmsy
\def \eightpoint{\def\rm{\fam0\eightrm}
\textfont0=\eightrm \scriptfont0=\sixrm \scriptscriptfont0=\fiverm
\textfont1=\eighti \scriptfont1=\sixi   \scriptscriptfont1=\fivei
\textfont2=\eightsy \scriptfont2=\sixsy   \scriptscriptfont2=\fivesy
\textfont3=\tenex \scriptfont3=\tenex   \scriptscriptfont3=\tenex
\textfont\itfam=\eightit  \def\it{\fam\itfam\eightit}%
\textfont\slfam=\eightsl  \def\sl{\fam\slfam\eightsl}%
\textfont\ttfam=\eighttt  \def\tt{\fam\ttfam\eighttt}%
\textfont\bffam=\eightbf  \scriptfont\bffam=\sixbf
\scriptscriptfont\bffam=\fivebf  \def\bf{\fam\bffam\eightbf}%
\tt \ttglue=.5em plus.25em minus.15em
\setbox\strutbox=\hbox{\vrule height7pt depth2pt width0pt}%
\normalbaselineskip=9pt
\let\sc=\sixrm  \let\big=\eightbig  \normalbaselines\rm
\textfont\truecmr=\eighttruecmr
\scriptfont\truecmr=\sixtruecmr
\scriptscriptfont\truecmr=\fivetruecmr
\textfont\truecmsy=\eighttruecmsy
\scriptfont\truecmsy=\sixtruecmsy
}\let\nota=\eightpoint}

\newfam\msbfam   
\newfam\truecmr  
\newfam\truecmsy 
\newskip\ttglue
\ifnum\tipi=0\TIPIO \else\ifnum\tipi=1 \TIPI\else \TIPITOT\fi\fi


%
\def\V#1{\vec#1}
\def\T#1{#1\kern-4pt\lower9pt\hbox{$\widetilde{}$}\kern4pt{}}

\let\io=\infty
\def\fra#1#2{{#1\over#2}}
\let\0=\noindent

\def\guida{\leaders\hbox to 1em{\hss.\hss}\hfill}
\def\tende#1{\vtop{\ialign{##\crcr\rightarrowfill\crcr
              \noalign{\kern-1pt\nointerlineskip}
              \hglue3.pt${\scriptstyle #1}$\hglue3.pt\crcr}}}
\def\otto{{\kern-1.truept\leftarrow\kern-5.truept\to\kern-1.truept}}


\def\etc{\hbox{\it etc}}

\def\ie{\hbox{\it i.e.\ }}

\def\fiat{{}}


\def\nn{{\V\n}}\def\oo{{\V\o}}
\def\nn{{\V\n}}
\def\NN{{\cal N}}\def\FF{{\cal F}}
\def\={{ \; \equiv \; }}

\def\pps{{\V\ps{\,}}}

\def\2{{1\over2}}

\def\st{\scriptscriptstyle}
\let\\=\noindent

\def\*{\vskip0.3truecm}

\fiat
\numsec=1\numfor=1
\def\hh{\V h}\def\pps{\V \psi}\def\equu(#1){{Eq.$\,$\equ(#1)}}
\footline={\ifnum\pageno=1
\rlap{\hbox{$\st 15\ marzo\ 1995$}}\hss\tenrm\folio\hss
\else
\hss
\folio\hss\fi}
\vglue2truecm
\centerline{\bf Lindstedt series and Kolmogorov theorem.}
\*\*
\centerline{\it Giovanni Gallavotti\footnote{${}^*$}{\nota%
Dipartimento di Fisica, Universit\`a di Roma, P. Moro 2, 00183 Roma,
Italia. Expanded version of a talk at the ``Congress on dynamical
systems'' Montevideo, Uruguay, march 27- april 1, 1995; archived in {\it
mp$\_$arc@ math. utexas. edu}, \#95-152 and in {\it chao-dyn@
xyz. lanl. gov}, \# 9503??, last version available by mosaic: {\it
http://chimera. roma1. infn. it}.}}
\*
{\it Abstract: the KAM theorem from a combinatorial viewpoint.}
\*\*

Lindstedt, Newcomb and Poincar\'e introduced a remarkable trigonometric
series motivated by the analysis of the three body problem, [P].  In
modern language, [G2], it is the generating function, that I call {\it
Lindstedt series} here, $\hh(\pps)=
\sum_{k=1}^\io \e^k \hh^{(k)}(\pps)$ of the sequence $\hh^{(k)}(\pps)$
of trigonometric polynomials associated with well known combinatorial
objects, namely {\it rooted trees}.

It is necessary to recall, first, the notion of rooted tree as used
here.  We lay down one after the other on a plane $k$ pairwise distinct
unit segments oriented from one endpoint to the other (respectively the
{\it initial point} and the {\it endpoint} of the oriented segment also
called {\it arrow} or {\it branch}).

The rule is that after laying down the first segment, the {\it root
branch}, with the endpoint at the origin and otherwise arbitrarily, the
others are laid down one after the other by attaching an endpoint of a
new branch to an initial point of an old one and leaving free the branch
initial point.  The set of initial points of the object thus constructed
will be called the set of the tree {\it nodes}.  A tree is therefore a
partially ordered set with top point the endpoint of the root branch,
{\it also} called the root (which is not a node).

The angles at which the segments are attached will be irrelevant: \ie
the operation of changing the angles between arrows emerging from the
same node (each arrow carrying along, unchanged, the subtree of arrows
possibly attached to its initial point) generates a group of
transformations and two trees that can be overlapped by acting on them
with a group element are regarded as identical.  The number of trees
with $k$ branches is thus bounded by $4^k k!$.

With each tree node $v$ we associate an {\it incoming momentum}, or
"decoration", which is simply an integer component vector $\nn_v$; with
the root of the tree (which is not regarded as a node) we associate a
label $j=1,\ldots,l$.

With each branch $\l=v'v$, with final point $v'$ and initial point $v$,
we associate another integer component vector, the {\it branch momentum}
"flowing through the branch", defined by $\nn(v)=\sum_{w\le v}\nn_w$ (we
shall also denote $\nn(v)$ by the symbol $\nn(\l)$).  Then, given a
positive matrix $J$ and a trigonometric polynomial
$f(\pps)=\sum_{0<|\nn|<N} f_\nn \cos\nn\cdot\pps$, $f_\nn=f_{-\nn}$, we
consider from now on only decorated trees $\th$ with $k$ branches, {\it
such that $\nn(v)\ne\V0$ for all $v$}, and associate with each decorated
tree the {\it value}:

$$V_f(\th)_j=-i\prod_{v<r} f_{\nn_v}\fra{\nn_{v'}\cdot J^{-1}
\nn_v}{(i\oo\cdot\nn(v))^2}
\Eq(1)$$

\0where $v'$ is the node immediately following $v$ in $\th$;
$\oo\cdot\nn(v)$ will be called the {\it divisor} of $v'v$; here $\nn_r$
denotes the unit vector in the $j$-th direction $\V e_j$,
$j=1,\ldots,l$. The momentum flowing through the root will be denoted
also $\nn(\th)$.  The {\it Lindstedt--Newcomb--Poincar\`e} ("LNP")
polynomial $\V h^{(k)}(\pps)$ is {\it defined} by $\sum_\nn \V
h^{(k)}_\nn e^{i\nn\cdot\pps}$ with:

$$\V h^{(k)}_{\nn,j}=\fra1{k!}\sum_{\th, \nn(\th)=\nn}
V_f(\th)_j\Eq(2)$$

\0The main result is, if $|\oo\cdot\nn|^{-1}< C|\nn|^\t$ for some
$C,\t>0$:
\*
\0{\bf Theorem:}{\it\ The Lindstedt series is convergent for $\e$ small
enough.}
\*
{\it Remark:} The first proof (of a more general result) is due to
Kolmogorov, [K]; a conceptually new proof came much later and is due to
Eliasson, [E], quite difficult to read but definitely {\it very nice and
correct}.  Further proofs were given by [G2] (extracted from [G1]) and
by [CF1]: somewhat different from the proof in [E] (and remarkably close
to each other (except for the size), although independent).  The main
contribution of the latter papers was to realize that the proof could be
greatly simplified by restricting it to a special case.  The above case
is the simplest and it is discussed in [G1],[G2]: even mild
generalizations of the simplest case, [CF1], tend to become quite
involved particularly if inappropriate notations are used: a comparative
analysis of the different proofs will appear in [GM4].
\*

{\it proof:} Let us call a branch $v'v$ of a tree $\th$ a {\it scale
$n$ branch}, if $2^{n-1}< C\,|\nn(v)\cdot\oo|\le 2^n$ where
$n=0,-1,-2,\ldots$ or a scale $1$ branch if $C\,|\oo\cdot\nn(v)|>1$.

We call a {\it cluster} of scale $n$ in the tree $\th$ a maximal
connected set of branches of scale $m\ge n$.

Notable clusters consist of the clusters $V$ with only one entering
branch $\l_V$ (and, of course, only one exiting branch $\l_V'$). If
$n_V$ is the scale of the cluster then the scale $n_{\l_V}$ of the
branch $\l_V$ is necessarily $n_{\l_V}< n_V$. If $v\in V$ call
$\nn^0(v)=\sum_{w\in V, w\le v}\nn_w$.

A cluster $V$ of the above type is called a {\it resonance} or a {\it
resonant cluster} if $\nn(\l_V)\=\nn(\l'_V)$ and {\it also} the number
$\NN$ of branches inside the cluster is not too large: $\NN<\fra1N
2^{-(n_{\l_V}+3)/\t}$; hence it is so small that
$|C\,\oo\cdot\nn^0(v)|>2^{n_{\l_V}+3}$.  A resonant cluster may contain
resonant subclusters: we shall denote $\tilde V$ the set of nodes in $V$
which are outside the possible resonant subclusters of $V$, and we call
them the {\it free nodes} of $V$.

A resonance has two scales attached to it: the scale $n_V$ of the
resonant cluster and the scale $n_{\l_V}$ of the branch entering the
resonance. We shall distinguish them by calling them the {\it inner
scale} and, respectively, the {\it outer scale} of the
resonance: it is of course $n_V>n_{\l_V}$.  But the latter inequality
can be made stronger: in fact the previous paragraph shows that the
scale $n_\l$ of a branch $\l$ in $V$ is determined by the size of the
``divisor'' $C\,(\oo\cdot\nn^0(\l)+\e)$ with $C|\e|<2^{n_{\l_V}}$, so
that the size of the divisor is $\ge 2^{n_{\l_V}+3}-2^{n_{\l_V}}>
2^{n_{\l_V}+2}$, hence:

$$n_{\l}\ge n_V\ge n_{\l_V}+3\Eq(3)$$
for all $\l$ in a resonance $V$ with outer scale $n_{\l_V}$ and inner
scale $n_V$.

Given a branch with endpoint on a node $v$ of a tree $\th$ we define the
operation of {\it attaching} or {\it shifting} the branch to another
node $w$: it consists in detaching {\it the whole subtree} preceding
$v$ (which has $v$ as root) and of attaching it to another node $w$: the
result is a new tree $\th'$.

One then considers all the trees $\th$ obtained from a given tree
$\th_0$ by one of the following operations:
\*
{\it%
\0(i) detaching the endpoint of a branch $\l_V$ entering a resonance
$V$ and attaching it (in the above sense) to all possible free nodes $v\in
\tilde V$: an act that will be called the {\it shift} operation on $V$.

\0(ii) changing the sign of all (simoultaneously) the incoming momenta
$\nn_v$ of the free nodes $v\in \tilde V$:
the {\it parity} operation on $V$.}

\*
\0This produces a family $\FF(\th_0)$ of trees containing not too many
trees: at most $\prod_{V} 2m_V$ if $m_V$ is the number of free nodes inside
a resonance $V$.  Since $\sum_V m_V\le k$ the number of elements of
$\FF(\th_0)$ does not exceed $\prod_V 2m_V\le2^{2k}$, (using $m\le 2^m$).

Furthermore a condition will be imposed on $\oo$ (see below) such that
given two trees $\th_0$ and $\th_1$ either $\FF(\th_0)=\FF(\th_1)$ or
$\FF(\th_0)\cap\FF(\th_1)=\emptyset$.  This can be done for the
following reason.

Consider a branch $\l_V$ with scale $n_{\l_V}$ entering a resonant
cluster $V$ with inner scale $n_V>n_{\l_V}$ and attached to a node
$v_1$ in $V$; then if $\l_V$ is shifted to be attached to a node $v_2\in
V$ the momenta flowing through the branches inside $V$ may change by an
amount $\nn(\l_V)$: therefore their scale may change by at most
$2^{n_{\l_V}}$.

If the resonance $V$ is inside many resonances $V_2,V_3,\ldots$ with
outer scales $n_2>n_3>..$ (note that $n_{\l_V}>n_2$) then by shifting
the branches $\l_{V_i}$ entering $V_i$ in all possible ways to be
attached at different free nodes inside $V_i$ the scale of the branches
in $V$ cannot change by more than $2^{n_{\l_V}}+2^{n_2}+2^{n_3}+\ldots<
2^{n_{\l_V}+1}$.

Since the number $\NN< \fra1N 2^{-(n_{\l_V}+3)/\t}$ of branches inside
the resonance $V$ is not too large the vector $\nn^0(v)=\sum_{w\in V,
w\le v}\nn_w$ is not longer than $2^{-(n_{\l_V}+3)/\t}$ for any $v\in
V$.  Hence assuming that the following property holds for $\oo$:

$$\min_{n\le p\le 0}| C |\oo\cdot\nn|-2^p|> 2^{n+1} \qquad
{\rm if}\quad n\le0,\ |\nn|<2^{-(n+3)/\t}\Eq(4)$$
we see that the number $x_v=C\oo\cdot\nn^0(v)$ is not only $>
2^{n_{\l_V}+3}$, but also it is further away from the extremes of
the interval $I_p=(2^{p-1},2^p]$ or, for $p=1$, $I_1=(1,+\io)$ containing
$x_v$, by an amount $2^{n_{\l_V}+1}$ at least.

This means that by shifting the branches entering the resonances external
to a resonant cluster $V$ of outer scale $n_{\l_V}$ one cannot change the
scales of the branches inside $V$ (because for $v\in V$ it is
determined by the size of $C \oo\cdot\nn^0(v)$ plus a quantity which is
at most $2^{n_{\l_V}+1}$, while the distance of $C\oo\cdot\nn^0(v)$
from the extremes of the intervals $I_p$ is greater).

{\it The scales of corresponding branches in the elements of a family
$\FF(\th_0)$ do not change by shifting operations nor they do change by
the parity operations}, provided \equu(4) holds: in particular
$2^{n_\l-1}<|C\oo\cdot\nn^0(\l)|\le 2^{n_\l}$, for any $\l$ in $V$. Here
two branches of two trees in $\FF(\th_0)$ are "corresponding" if they
can be superposed by performing suitable sequences of shift or parity
transformations in (i), (ii) above. The part of the statement concerning
the parity is easily checked, too.

Therefore, assuming \equu(4) for the time being (note, however, that
almost all $\oo$'s verify \equu(4)) we shall use the following
representation of the function $\V h^{(k)}(\pps)$: \* \0{\it Lindstedt
series representation:

(a) consider all the trees with $k$ branches and collect them into
disjoint families: each family is obtained from a tree $\th$ in it by
the operation of shifting the branches entering a resonance $V$ to the
various free nodes in $V$, or of reversing simoultaneously the signs
of the momenta $\nn_v$ incoming into the free nodes of $V$,
for all resonances $V$, independently.

(b) define the {\it value of a family} to be the sum of the values of
the trees in it.

(c) sum all the values of the families.}
\*
This is a natural representation for $\V h^{(k)}$: it was introduced
in [G1],[G2] and with a minor variation, based on the slightly
different notion of resonance used by [E], in [CF1] (independently).

The above representation seems also close to the one proposed in [GL].
In fact it is probably very natural if one recalls that there is a
symplectic structure in the problem that generates the Lindstedt series,
see comments (7,8,9) below.  Evidence in this direction is provided by
the alternative scheme of proof followed by [E], where the bound on the
convergence radius of the series is deduced from an argument in some
respect simpler (but abstract) than the one below, where the
cancellations happening when one sums all the tree values of trees in a
given family are checked "indirectly" as a consequence of the symplectic
symmetry of the mechanical system whose theory leads to the series,
through an idea that goes back to [P], see [GM4].

The proof will now proceed to the easy check that the value of each
family is bounded by $B^k$ for some $B>0$: since there are at most as
many families as trees (\ie $\le k! 4^k$) this implies the theorem.

\def\VV{{\cal V}}
Fix a family $\FF(\th)$ of trees.  And consider the set of the
innermost, or {\it first order}, resonances $\VV_1$.  Then we shall
call $\e_V=C \oo\cdot\nn(\l_V)$ where $V\in\VV_1$ is a resonant cluster
with incoming momentum $\nn(\l_V)$ and we shall sum over the
trees obtained from $\th$ by the shift and
parity operations on the resonances in $\VV_1$: if $2m_1{\buildrel def
\over =}\sum_{V\in\VV_1} 2m_V$ the number of trees involved in the sum
is $\prod_{V\in\VV_1} 2m_V\le 2^{2m_1}$, (by $2m< 2^{2m}$).

The "value $F_1(\th)$ of the subfamily" of $\FF(\th)$ so obtained {\it
can be regarded} as a function of the complex parameters $\{\e_V\}$,
with $V$ varying in $\VV_1$ defined as follows.  Note that the momentum
flowing in a branch $\l=v'v\subset V$ is
$\oo\cdot\nn^0(v)+\s_v\,\oo\cdot\nn(\l_V)$ with $\s_v=0,1$; then the
function is constructed by replacing the quantities
$(\oo\cdot\nn^0(v)+\s_v\,\oo\cdot\nn(\l_V))^{-2}$ by
$(\oo\cdot\nn^0(v)\l+\s_v\e_V)^{-2}$ for all the branches $v'v\subset
V$.

Note that this implies that in regarding the value of $\FF_1(\th)$ as a
function of the $\e_V$'s the divisors associated with the branches
entering or exiting the resonance $V$ {\it are not} represented as
$\e_V^{-2}$.

It is now easy to see that the value of a family is holomorphic in the
variables $\e_V$ in the disks $|\e_V|<2^{n_{V}-3}$.  This is because in
any tree of the family it is $|C\,\oo\cdot\nn^0(v)|>2^{n_{V}-1}$ (by
the remarks following \equu(4)) so that for $|\e_V|<2^{n_{V}-3}$ it is
$|C\,\oo\cdot\nn^0(v)+\s\e_V|>2^{n_{V}-3}$.  In fact, for later use,
one sees that the last inequality holds even in the larger disk
$|\e_V|<2^{n_V-3}(1+\fra17)$ (the $\fra17$ will arise later as the sum
of the geometric series with ratio $\fra18$).

This means that the value $F_1(\{\e_V\}_{V\in\VV_1})$ of the subfamily
$\FF_1(\th)$ can be bounded above, for $|\e_V|<2^{n_{V}-3}$ (and in
fact even for $|\e_V|<2^{n_{V}-3}\fra87$), by trivially bounding the
divisors in the contributions of the various trees in terms of their
scales, or by using $|C\,\oo\cdot\nn^0(v)+\s\e_V|\ge2^{n_{V}-3}$.  One
finds therefore, from \equu(1), the bound:

$$2^{2m_1}X{\buildrel def\over =}
2^{2m_1}\fra{1}{k!} \big(\fra{f_0 C^2 N^2}{J_0}\big)^k
\, \prod_\l 2^{-2(n_\l-3)}=2^{2m_1}\big(\fra{f_02^{6}C^2}{J_0}\big)^k
\,\prod_n 2^{-2n N_n} \Eq(5)$$
if $N_n$ is the number of branches with scale $n$, $f_0=\max
|f_\nn|$ and $J_0$ is a lower bound to $J$.

It is easy to check that the function $F_1(\{\e_V\})$ vanishes in {\it
each} $\e_V$ to {\it second order} for $\e_V=0$.  This is simply
because the sum of the momenta $\nn_{v_j}$ incoming into the nodes
${v_j}\in \tilde V$, $\sum_j \nn_{v_j}$, in each resonance $V$,
vanishes, (by definition of resonance).  Furthermore the only variation
of the value of a tree when the branch entering a resonance is shifted
to the node ${v_j}$ from the node $v_{j'}$ is the replacement of a
factor $\nn_{v_{j'}}\cdot J^{-1}\nn'$, for some $\nn'$, with the factor
$\nn_{v_{j}}\cdot J^{-1}\nn'$, when $\e_V=0$.  Hence if, $\e_V=0$ for
some $V$,  $F_1(\{\e_V\})$ is proportional to $\sum_{v\in \tilde
V}\nn_v=\V0$ and vanishes to first order in each of the $\e_V$.  The
summation over the sign reversals turns the function $F_1(\{\e_V\})$
into an even function of the $\e_V$, hence vanishing to second order.

Therefore the above bound \equu(5) in the domain $|\e_V|< 2^{n_{V}-3}$
implies, by the maximum principle, that the bound can be improved
by a factor
$\prod_{V\in \VV_1} \big(\fra{|\e_V|}{2^{n_{V}-3}}\big)^2$; which
becomes, if one replaces $\e_V$ by its actual value:

$$Y_1{\buildrel def\over =}\prod_{V\in\VV_1}
2^{2(n_{\l_V}-n_V+3)}\qquad {\rm if}\quad n_{\l_V}\le n_V-3\Eq(6)$$
Note that only the case $n_{\l_V}< n_V-3$ needs the maximum priciple.

We now consider the {\it second order resonances}, \ie the set of
resonances $\VV_2$ which only contain first order resonances.  Then we
denote again $\e_V=C \oo\cdot\nn(\l_V)$ where $V\in\VV_2$ is a resonant
cluster with incoming momentum $\nn(\l_V)$ and we shall set
$2m_2=\sum_{V\in \VV_2}2m_V$, if $m_V$ is the number of free nodes in
$V$.

Summing over the $<2^{2m_1+2m_2}$ trees obtained from $\th$ by the
shift and parity operations on the resonances in $\VV_1,\VV_2$ we
define the "value $F_2(\th)$ of the subfamily" of $\FF_2(\th)$ so
obtained.  It can be regarded as a function of the complex parameters
$\{\e_V\}$, with $V$ varying in $\VV_2$.  The momentum flowing in a
branch $\l=v'v\subset V$ is $\oo\cdot\nn^0(v)+\s_v\oo\cdot\nn(\l_V)$
with $\s_v=0,1$ and the function is constructed, as in the previous
case, by replacing the quantities
$(\oo\cdot\nn^0(v)+\s_v\oo\cdot\nn(\l_V))^{-2}$ by
$(\oo\cdot\nn^0(v)\l+\s_v\e_V)^{-2}$ for all $v'v\subset V$.

Furthermore when the variable $\e_V$ varies in the disk $|\e_V|<
2^{n_V-3}$ the divisors of the branches entering the first order
resonances $W\subset V$ have the form
$\oo\cdot\nn^0(\l)+\s_v\oo\cdot\nn(\l_V)$ with $2^{n_{\l}-1}<
C|\oo\cdot\nn^0(v)|\le 2^{n_{\l}}$ and therefore if $n_V<n_W-3$ they are
bounded above by $2^{n_V}+2^{n_{V}-3}\le 2^{n_W-4}+2^{n_W-6}<
2^{n_W-3}$.

Hence if $n_V<n_W-3$ the value of the function is bounded by
$2^{2m_2+2m_1}Y_1 X$, for $|\e_V|< 2^{n_V-3}$: this remains true even if
$n_V=n_W-3$ for one or more $W\subset V$.  In the latter cases, in fact,
the momentum entering the resonance $W$ needs not be $< 2^{n_W-3}$ but
it is $\le 2^{n_V}+2^{n_V-3}=2^{n_W-3}+2^{n_W-6}<2^{n_V-3}\fra87$ so
that the branches $\l$ inside $W$ are still such that
$C|\oo\cdot\nn^0(\l)+\s_v \e_W|> 2^{n_W-1}-2^{n_W-3}-2^{n_W-6}\ge
2^{n_W-3}$ and taking $Y_1=1$ provides a correct bound.

Again the value $F_2(\th)$ vanishes to second order in the variables
$\e_V$ so that it can be bounded, if one replaces $\e_V$ by its actual
value, by $2^{m_2+m_1}Y_1Y_2 X$ where $Y_2$ is defined as in \equu(6)
with $\VV_2$ replacing $\VV_1$, \ie
$\prod_{V\in\VV_2}2^{2(n_{\l_V}-n_{V}+3)}$.  One now considers the
third order resonances and so on until one finds the bound on the value
of the complete family $\FF(\th)$ given by
$2^{2m_1+2m_2+\ldots}Y_1Y_2\ldots X$.  Since $2m_1+\ldots\le 2k$ one
finds, recalling the definition of $X$ in \equu(5):
$$\fra1{k!}\big(\fra{f_02^{6}C^2}{J_0}\big)^k\prod_n 2^{-2n N_n}
\prod_{n\le 0} \prod_{T,n_T=n}\prod_{i=1}^{m_T} 2^{2(n-n_i+3)}\Eq(7)$$
if $T$ are the clusters of the family of trees under consideration,
$m_T$ is the number of maximal resonances contained inside the cluster
$T$, and $n_i$ denotes the scale of the branch entering the $i$-th of
such resonances.

The number $N_n$ of branches of scale $n$ can be bounded by imitating
the key bound due to Siegel and Brjuno in the case of trees without
resonances (the bound seems to have been well known since Siegel's
theorem for the similar case of the linearization of the complex maps,
which essentially corresponds to the analysis of the generating function
of polynomials like the ones considered here and generated by
resonanceless trees).

If $m_T$ is the number of maximal resonances inside the cluster $T$
then $N_n$ is bounded by:

$$N_n\le 4 N 2^{(n+3)/\t}k+\sum_{T,n_T=n} (m_T-1)\Eq(8)$$
the proof in [G2] is reproduced, for completeness, in the appendix.

Hence the product in \equu(7) is bounded by combining \equu(7) and \equu(8),
after some simple power counting, by:
$$\fra1{k!}\big(\fra{f_02^{12}C^2N^2}{J_0}\big)^k\prod_{n=-\io}^1
2^{-8nN 2^{(n+3)/\t}k}=\fra1{k!} B_0^k\Eq(9)$$
with $B_0= C^2 f_0 J_0^{-1}N^2 2^{12}2^{-\sum_n 8n N2^{(n+3)/\t}}$.

The number of families cannot exceed the number of trees, \ie $k!
2^{2k}$, so that the series \equu(1) converges for $|\e|<(4B_0)^{-1}$,
which is even a rather good estimate, at least as a first estimate, for
$N$ small.

This completes the proof in the case in which the inequalities \equu(4)
hold. Such inequalities were called {\it strong diophantine property} in
[G1], [G2]. If one examines carefully their role (\ie keeping the scales
unchanged while performing the operations (i), (ii)) one sees that it is
very reasonable that they can be replaced by a suitable similar property
which holds in the general case in which only the diophantine inequality
(following \equu(2) above) holds: this is discussed at the end of [G2]
and proved in [GG].
\*
{\it Comments:}
\*
(1) The restriction that $f$ is a trigonometric polynomial can be lifted
easily to treat analytic $f$'s, as shown in [GM2], (in [CF1] too the $f$
is not supposed a polynomial).  The restriction that $f$ is even can be
lifted simply by replacing the parity operation by the {\it new}
operation of shifting the node to which the {\it outgoing branches} are
attached in $\tilde V$, as shown in [CF1].  Note that the even case is
extremely interesting for the applications as the perturbation functions
$f$, in many cases, are even because this property is related to the
time reversal symmetry (see [BCG]).

(2) The above method can be extended to study the generating functions
of quantites originated in a way analogous to the Lindstedt polynomials
in the perturbation theory of whiskered tori, see [G1]: the first such
results are in [Ge1], [Ge2]. In the later review paper [CF2], aware of
such references, a part of such results are also discussed and
apparently claimed as new, if so improperly because the analysis in
[Ge1],[Ge2] is correct and complete (and, furthermore, it goes quite far
beyond what is discussed in [CF2] in the cases common to both papers).

(3) The method of [E] is not quite the same as the above, see [GM4].

(4) The above method reminds very much of field theory methods:
the similarity was noted first in [FT], at a formal level, without
taking advantage of it to study a proof of the theorem. The results of
[GM1], [GM2] take full advantage of the similarity and build a proof
(see comment (9) below).

(5) In [G3] it is shown that there is a euclidean field theory whose one
point Schwinger function is the function \equu(2). This analogy  is
pushed further in [GGM], where a connection with the theory of the
singularity in $\e$ of the function \equu(1) as $\e$ grows is
heuristically attempted, along an early suggestion in [PV].

(6) the convergence of the Lindstedt series generated by the general
KAM theorem can be proved along the above lines, and as one should expect,
it does not involve any new ideas, in fact see [CF2] for a somewhat more
detailed sketch and [GM3] for a proof.

(7) there are other ways besides the one of [E] to reach a
representation similar to the above and giving immediately an absolutely
convergent series for the generating function of the LNP polynomial.
This has been shown in [GL] and it seems closely related to the fact
that the problem is associated with the theory of hamiltonian
evolutions. Also the method of [E] does not perform explicit
cancellations on individual terms of the series, but it shows that they
must happen because of the symplectic nature of the equations that the
convergence of the LNP series would solve; hence it would be interesting
to establish a clear connection between the method in [GL] and that of
[E]: this may help establishing a more clear connection also between the
above method and [G1],[G2] or [CF1].

(8) one can ask whether the above proof can be used for numerical
purposes.  At first sight one may think that the maximum principle used
in the proof is an obstacle.  However one should note that the sum over
the tree values in a given family is an algebraic operation: hence the
analytic estimate can be replaced by an algebraic one (this is
exploited in [GM1], [GM2]).

\0When one performs the sum corresponding to the shift of the incoming
branch, or to the parity transformation, for a resonance $V$, one takes a
linear combination of $m_V-1$ products of divisors.  Therefore the
result is a polynomial in $\e_V$.  The above remarks show that it
starts at second order: hence one can factor $\e_V^2$ ({\it by a finite
number of algebraic operations}) which cancels one of the two divisors
associated with the resonance incoming and outgoing branches.

\0This means that each resonance only contributes a divisor not exceeding
$\e^{-2}_V 2^{-2n_V}\le$ $2^{-2n_{\l_V}-2n_V}$, in the estimates, {\it
instead of $\e_V^{-4}\le 2^{-4n_{\l_V}}$}: in other words the "extra"
divisor on scale $n_{\l_V}$ is replaced by a divisor on the generally
much higher scale $n_V$ (hence much larger).  It is not difficult to see
that the phenomenon of accumulation of divisors expressed by the sum in
\equu(8) is precisely due to the fact that without taking into account
the above remark the resonances contribute {\it two} divisors on the
same scale (\ie they generate a divisor $\e_V^{-4}$).  Hence once the
above algebraic operation is performed one has no more divisors in
excess, in comparison with the Siegel's situation: this idea could also
be used to provide a quick proof of \equu(8) if one assumes it to be
valid for resonanceless trees, a case in which it was well known being
implicit in Siegel's and Brjuno's work.

\0The remarkable and important result in [GL], that justifies the
apparent absence of a cancellation analysis, is that the above
collection of terms is automatically performed if one proceeds by keeping
always explicitly into account the symplectic structure of the problem
that is solved by the Lindstedt series.  In this respect the work [GL]
is much closer to [E] than [G1],[G2],[CF1].

(9) of course one could define the notion of resonant cluster by
forgetting the condition on the number of branches: this would lead, if
suitably complemented with a redefinition of the families of trees, to
an {\it overcompensation} and to worse final estimates (and to a
trivially more involved proof).  This is done in [E], and in some way
also in [GL] (but not in [G1],[G2],[CF1]).  On the other hand it can
provide a conceptual simplification as it has been shown in [GM2],[GM3]
where the possibility of overcompensating is even exploited to use
deeply the analogies of the above ideas with those of renormalization
theory in quantum field theory (see (4) above). The result is what I
think is the most transparent and elementary proof of the KAM theorem,
even simpler than the one decribed in the present paper (although the
latter is likely to be better for numerical applications).

(10) finally it is important to note that the results depend only on
the smallest eigenvalue $J_0$ of the matrix $J$ and {\it not} on the
maximum.  Given the hamiltonian
$H=\oo\cdot\V A+\fra12\V A J^{-1}\V A+\e\f(\V\a)$ where $\V A\in R^l$ and
$\V\a$ is in the torus $T^l$ ({\it Thirring model}, see [G2]),
the Lindstedt series solves, see [G2],
the problem of the existence of a KAM torus close to $\V A=\V0$ and
with rotation vector $\oo$: hence we see that the condition on the size
of $\e$ is {\it independent} on the {\it twist rate}, which is the {\it
maximum} eigenvalue of $J$.  In the latter model the twist condition,
that plays such an important role in the general KAM theorem, is {\it
not necessary}: thus tori arising from the above analysis were
called, in [G1],[G2], {\it twistless KAM tori}.  \*

\penalty-200
{\bf Appendix A1: Resonant Siegel-Brjuno bound.}
\penalty10000\*

Calling $N^*_n$ the number of branches of scale $\le n$ which are {\it
not} entering a resonance ("non resonant branches").  We shall prove
first that $N^*_n\le 2k (N 2^{-(n+3)/\t})^{-1}-1$ if $N_n>0$.  We fix
$n$ and denote $N^*_n$ as $N^*(\th)$.  Let $\e=\fra1\t$ and
$E=N^{-1}2^{-3\e}$.

If the root branck scale of $\th$ is $>n$ and if
$\th_1,\th_2,\ldots,\th_m$ are the subtrees of $\th$ attached to the
root branch of $\th$ and with $k_j>E\,2^{-\e n}$ branches, it is
$N^*(\th)=N^*(\th_1)+\ldots+N^*(\th_m)$ and the statement is inductively
implied from its validity for $k'<k$ provided it is true that
$N^*(\th)=0$ if $k<E2^{-\e n}$, which is is certainly the case
if $E$ is
chosen as above.\footnote{${}^1$}{\nota Note that if $k\le E\,2^{-n\e}$
it is, for all momenta $\nn$ of the branches, $|\nn|\le N E \,2^{-n\e}$,
\ie $|\oo\cdot\nn|\ge(NE\,2^{-n\e})^{-\t}=2^3\,2^{n}$ so
that there are {\it no} clusters $T$ with inner scale $n_T=n$ and
$N^*=0$.}

In the other case it is $N^*_n\le 1+\sum_{i=1}^mN^*(\th_i)$, and if
$m=0$ the statement is trivial, or if $m\ge2$ the statement is again
inductively implied by its validity for $k'<k$.

If $m=1$ we once more have a trivial case unless the order $k_1$ of
$\th_1$ is $k_1>k-\fra12 E\,2^{-n\e}$.  Finally, and this is the real
problem as the analysis of a few examples shows, we claim that in the
latter case the root line of $\th_1$ is either a resonant line or it has
scale $>n$.

Accepting the last statement it will be: $N^*(\th)=1+N^*(\th_1)=
1+N^*(\th'_1)+\ldots+N^*(\th'_{m'})$, with $\th'_j$ being the $m'$
subdiagrams emerging from the first node of $\th'_1$ with orders
$k'_j>E\,2^{-\e n}$: this is so because the root line of $\th_1$ will
not contribute its unit to $N^*(\th_1)$.  Going once more through the
analysis the only non trivial case is if $m'=1$ and in that case
$N^*(\th'_1)=N^*(\th"_1)+\ldots+N^*(\th"_{m"})$, \etc, until we reach a
trivial case or a diagram of order $\le k-\fra12 E\,2^{-n\e}$.

It remains to check that if $k_1>k-\fra12E\,2^{-n\e}$ then the root
branch of $\th_1$ has scale $>n$, unless it is entering a resonance.

Suppose that the root line of $\th_1$ has scale $\le n$ and is not
entering a resonance.  Note that
$|\oo\cdot\nn(v_0)|\le\,2^n,|\oo\cdot\nn(v_1)|\le \,2^n$, if $v_0,v_1$
are the first vertices of $\th$ and $\th_1$ respectively.  Hence
$\d\=|(\oo\cdot(\nn(v_0)-\nn(v_1))|\le2\,2^n$ and the diophantine
assumption implies that $|\nn(v_0)-\nn(v_1)|> (2\,2^n)^{-\t^{-1}}$, or
$\nn(v_0)=\nn(v_1)$.  The latter case being discarded as
$k-k_1<\fra12E\,2^{-n\e}$ (and we are not considering the resonances:
note also that in such case the lines in $\th/\th_1$ different from the
root of $\th$ must be inside a cluster, see footnote 3)), it follows
that $k-k_1<\fra12E\,2^{-n\e}$ is inconsistent: it would in fact imply
that $\nn(v_0)-\nn(v_1)$ is a sum of $k-k_1$ vertex modes and therefore
$|\nn(v_0)-\nn(v_1)|< \fra12NE\,2^{-n\e}$ hence $\d>2^3\,2^n$ which is
contradictory with the above opposite inequality.

The total number of branches with scale $n$ is therefore $N^*_n+
\sum_{T, n_T=n} m_T$ if $m_T$ is the number of resonances contained in
$T$.  A similar, far easier, induction can be used to prove that if
$N^*_n>0$ then the number $p$ of clusters of scale $n$ verifies the
bound $p\le 2k \,(E2^{-\e n})^{-1}-1$; it will be left out (see [GG]).
Thus \equ(8) is proved.

{\it Remark}: the above argument is a minor adaptation of Brjuno's
proof of Siegel's theorem, as remarkably exposed by P\"oschel, [P\"o].

\*
{\it Acknowledgements:} I am indebted to G.  Gentile and V.
Mastropietro for many discussions and for explaining important details
of the work of Eliasson, that I had missed, and for clarifying comments
on his work as well as on their own work. I am grateful to G. Gentile
for his careful comments on the details of the paper. This work is part
of the research program of the European Network on: "Stability and
Universality in Classical Mechanics", \# ERBCHRXCT940460.
\*\*
\0{\bf References}
\*
\0[BCG] Benettin, G., Carati, A., Gallavotti, G.: {\it
A rigorous implemenation of the Jeans--Landau--Teller approximation for
adiabatic invariants}, archived in {\it mp$\_$arc@ math. utexas. edu},
\# 95--???.

\0[CF1] Chierchia, L., Falcolini, C.: {\it  direct proof of a theorem by
Kolmogorov in hamiltonian systems}, Annali della Scuola Normale
Superiore di Pisa, {\bf21}, 541--593, 1994.

\0[CF2] Chierchia, L., Falcolini, C.: {\it
Compensations in small divisor problems}, Archived in {\it mp$\_$arc@
math.  utexas.  edu}, \#94-270.

\0[E] Eliasson L. H.: {\it Hamiltonian systems with linear normal
form near an invariant torus}, ed. G. Turchetti, Bologna Conference,
30/5 to 3/6 1988, World Scientific, 1989. And {\it Generalization of an
estimate of small divisors by Siegel}, ed. E. Zehnder, P. Rabinowitz,
book in honor of J. Moser, Academic press, 1990. But mainly:
{\it Absolutely convergent series expansions
for quasi--periodic motions}, report 2--88, Dept. of Math., University of
Stockholm, 1988.

\0[FT] Feldman, J., Trubowitz, E.: {\it Renormalization
in classical mechanics and many body quantum field theory}, Journal
d'Analyse Math\'ematique, {\bf 58}, 213-247, 1992.

\0[G1] Gallavotti, G.: {\it Twistless KAM tori, quasi flat
homoclinic intersections, and other cancellations in the perturbation
series of certain completely integrable hamiltonian systems.  A
review.},  Reviews on Mathematical Physics, {\bf 6}, 343-- 411, 1994.

\0[G2] Gallavotti, G.: {\it Twistless KAM tori}, Communications in
Mathematical Physics, {\bf 164}, 145--156, 1994.

\0[G3] Gallavotti, G.: {\it Perturbation Theory}, in "Mathematical
physics towards the XXI century", p.  275--294, ed.  R.  Sen, A.
Gersten, Ben Gurion University Press, Ber Sheva, 1994.

\0[GG] Gallavotti, G., Gentile, G.: {\it Non recursive proof of the KAM
theorem}, preprint, p.  1-8, agosto, 1993, Roma.  Archived in
{\it mp$\_$arc@ math.  utexas.  edu}, \#93-229,
in print on Ergodic Theory and Dynamical Systems with the title
{\it Majorant series convergence for twistless KAM tori.}

\0[Ge1] Gentile, G.: {\it A proof of existence of whiskered tori with
quasi flat homoclinic intersections in a class of almost integrable
systems}, preprint Roma ``La Sapienza" n. 996, february 1994, to appear
in Forum Mathematicum. archived in {\it mp$\_$arc@ math.  utexas.  edu},
\#94-33, in print on Forum Mathematicum.

\0[Ge2] Gentile, G.: {\it Whiskered tori with prefixed
frequencies and Lyapunov spectrum}, preprint Roma ``La Sapienza" n.
1015, april 1994. archived in {\it mp$\_$arc@ math.  utexas.  edu},
\#94-92.

\0[GL] Giorgilli, A., Locatelli, U.: {\it Kolmogorov theorem and
classical perturbation theory}, preprint Dipartimento Matematica,
Milano, january 1995.

\0[GM1] Gentile, G., Mastropietro, V.: {\it KAM theorem revisited},
archived in {\it mp$\_$arc@ math. utexas. edu}, \#94-403.

\0[GM2] Gentile, G., Mastropietro, V.: {\it Tree expansion and multiscale
analysis for KAM tori}, archived in {\it mp$\_$arc@ math. utexas. edu},
\#94-405.

\0[GM3] Gentile, G., Mastropietro, V.: in preparation,
preprint, 1995.

\0[GM4] Gentile, G., Mastropietro, V.: in preparation,
preprint, 1995.

\0[GGM] Gallavotti, G., Gentile, G., Mastropietro, V.: {\it Field
theory and KAM theorem}, Archived in {\it mp$\_$arc@ math.  utexas.
edu}, \#95-???.

\0[K] Kolmogorov, N.: {\it On the preservation of conditionally
periodic motions}, Doklady Aka\-de\-mia Nauk SSSR, {\bf 96}, 527-- 530,
1954.  See also: Benettin, G., Galgani, L., Giorgilli, A., Strelcyn,
J.M.: {\it A proof of Kolmogorov theorem on invariant tori using
canonical transormations defined by the Lie method}, Nuovo Cimento,
{\bf 79 B}, 201-- 223, 1984.

\0[P\"o] P\"oschel, J.: {\it Invariant manifolds of complex analytic
mappings}, Les Houches, XLIII (1984), vol. II, p. 949-- 964, Ed. K.
Osterwalder, R. Stora, North Holland, 1986.

\0[PV] Percival, I., Vivaldi, F.: {\it Critical dynamics and
diagrams}, Physica D, {\bf 33}, 304-- 313, 1988.

\0[P] Poincar\'e, H. {\it Les m\'ethodes nouvelles de la m\'ecanique
c\'eleste}, vol. II, ch. 9, 1893.

\bye